# Title

Machine Learning Guided Multiscale Design of DNA-functionalized Nanoparticles for Targeted Self-Assembly of the Double Gyroid

# Authors

Luis Nieves-Rosado[1] and Fernando Escobedo[1]

[1]Cornell University

# Abstract

In soft matter science, it is often the goal to design new materials with targeted properties. These materials can be used in many applications, each requiring specific features to be optimized for maximum fitness. The use of self-assembly processes, where a target structure is attained from suitably designed building blocks, is a powerful tool for this optimization. However, it is still an open question how to generally design realizable building blocks that lead to the desired phases. In this work, the double gyroid is chosen as target structure and DNA functionalized nanoparticles are used as our building blocks. Using existing pair potentials as inspiration, a large design space is defined for exploration of the target structure. An effective search strategy is implemented, where free energy calculations are first used to coarse grain our originally fine-grained model of the building blocks and then quickly evaluate the fitness of each design. These data are then fed into a machine learning algorithm that allows obtaining predictions for all candidates in our design space through an active learning loop. Successful coarse-grained designs are identified and evaluated again through interfacial pinning calculations with the fine-grained model. This work leads to the development of specific, experimentally relevant designs of DNA functionalized nanoparticles that self-assembly into the target phase. The methodology used can be extended to other types of building blocks and target structures.

## Introduction

The design of new soft materials with targeted properties has been a focus of the soft matter field for decades [1]. These materials can find applications in photonics [2-3], biomedical applications [4-5], catalysis [6] and other fields, with each requiring a unique combination of targeted features. To manufacture these materials, one must find the correct building blocks which, when properly designed, can undergo either spontaneous self-assembly or directed assembly into the desired structures.

How to connect a target structure to the correct building blocks in self-assembling processes is still largely an open question. This "inverse design" problem requires the optimization of the key features of the available building blocks, until the overall sought-after properties for the material are achieved. This problem has been tackled by powerful computational approaches [7-8], most recently with assistance from machine learning methods [9-10].

While some inverse methods can be applied directly to experiments [11], the high cost and slow timescales involved with repeated experimental iteration can make this process untenable. There is then a desire for ways to connect the results of inverse design methods, which often come in the form of pair potentials [12-14], to recipes that can directly be connected to experimental synthesis and fabrication.

One widely used platform for the design of particles with controllable interparticle interactions and assembling properties is that of DNA functionalized nanoparticles [15]. These particles, usually made of a gold core, have DNA strands attached to them, with a single stranded DNA "sticky" segment at the end. Changing the design of this single stranded segment allows for a tunable control of interparticle interactions via the degree of hybridization of complementary DNA strands attached to different particles. They thus provide a versatile building block for the assembly of many structures. While significant work has been done in studying the connection between the specific design of particles with the resulting structures [16-18] there remains a large parameter space of designs to be explored.

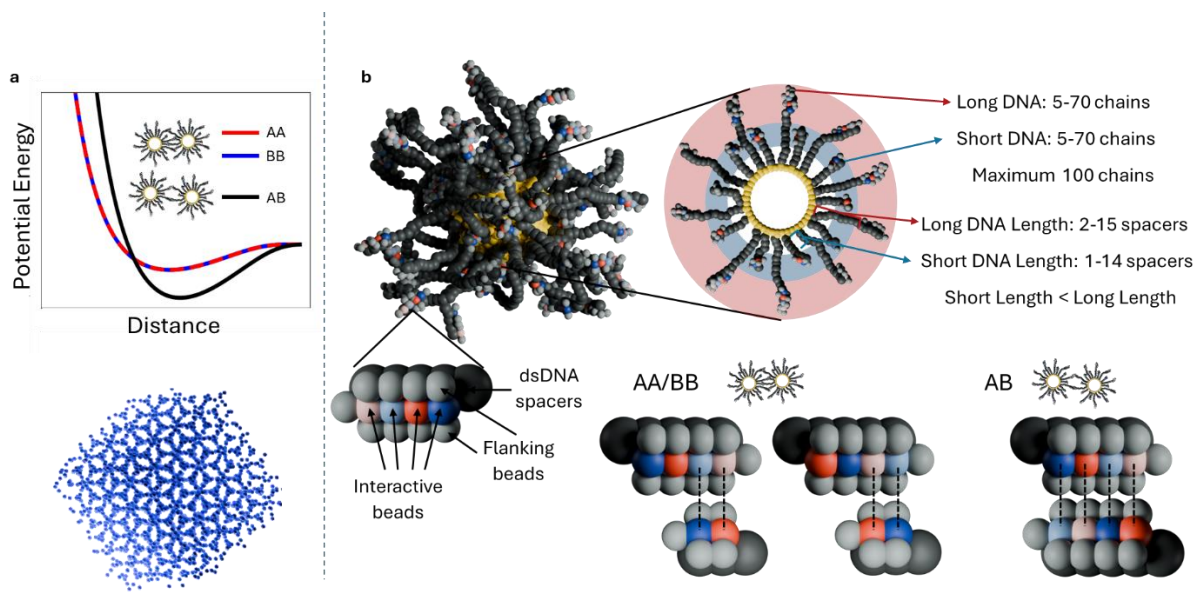

Figure 1: Target structure and design space of exploration. a. Our target structure is the double gyroid, shown in blue, as formed by the Kumar-Molinero interparticle potential [22], shown here for $\epsilon_{AB}/\epsilon_{AA/BB} = 2.1$ and $\sigma_{AB}/\sigma_{AA/BB} = 1.045$. b. 'Fine grained' description of DNA functionalized nanoparticles from Knorowski et al. [26], depicting the addition of both long and short DNA segments, which allow for the creation of two interaction layers which are calibrated to manipulate the strength and distance of AA/BB/AB bond pairs.

In this study we propose a multiscale framework that leverages machine learning methods to search the vast parameter space of DNA functionalized nanoparticles quickly and effectively by combining free energy coarse-graining methods with active learning optimization. As a goal design, we select the double gyroid phase, a networked structure which often appears in block copolymer systems [19] but is nontrivial to obtain with nanoparticle-based building blocks. While a similar phase has been obtained through the use of DNA functionalized nanoparticles in the past [20], a different design approach was used (particles with size asymmetry) which did not converge to the traditional double gyroid.

## Methodology

**Target Structure: Double Gyroid and the Kumar Molinero Potential**

We have chosen the double gyroid phase as the target structure. This phase has many emerging applications in optics and catalysis and plays a role in generating structure-based color in various insects, thanks in part to its highly regular, 3 dimensional network geometry. While challenging, it is a suitable choice to test assembly control of DNA functionalized nanoparticles, as it has a highly non-trivial structure which is difficult to obtain in nanoparticle-based systems [21]. Opportunely, there does exist in the literature a particle pair potential that self-assembles into this phase, which can provide some guidance for the design of our nanoparticles [22].

The Kumar-Molinero (KM) potential, shown in Figure 1-a, is a binary pair-wise additive potential, where particles of type A and B interact through the soft Stillinger-Weber potential [23]. This potential is calibrated such that the attraction of the AA and BB pairs is weaker than that of the AB pair but with the unusual property that the AA and AB pairs have a smaller equilibrium "contact" distance than that of the AB pair. This property induces local positive non-additivity, where the AB bond's actual length is greater than the sum of the radii of its components. This is a feature that has been shown to lead to interesting self-assembly [24].

This property, a stronger but longer AB contact, relative to the like contacts, is non-trivial to implement experimentally. The challenge of creating a particle design that has the KM potential features is similar to that faced when inverse designed potentials. However, as explored in Kumar and Molinero's original work [22], it is a key factor that allows the formation of a wide range of self-assembled phases analogous to those found in block-copolymers. Among them is the double gyroid phase in 33%-67% A-B mixtures, where the minority component forms the networks. See Figure 1-a for an example structure, showing only the minority component.

**'Fine grained' model for DNA functionalized nanoparticles**

While the KM potential provides the desired self-assembly behavior, there is no obvious recipe for how to design an actual particle that would interact in that manner. We propose the use of DNA functionalized nanoparticles to approach these interactions by employing multiple interaction layers, as shown in Figure 1-b. Our proposed design has double stranded DNA strands of two different lengths, with specific single stranded ends that control which inter-particle layers interact preferentially. Conceptually, the introduction of two length scales for the attraction wells allows for more flexibility in shaping up the pair potential function. Here, the AB interactions happen through hybridization of the long chains exclusively and occur through four (coarse grained) DNA attractive sites. The AA and AB interactions are designed to occur at closer distances, since hybridization happens through the long chain of one particle interacting with the short chain of another. To guarantee that the AB interaction is enthalpically favorable the short chains have fewer attractive sites.

It is important to note that the idea of using multiple length scales to control DNA functionalized nanoparticle assembly is not entirely new. Previous studies have used neutral polymer layers to introduce steric repulsion between nanoparticles to obtain enhanced crystallization [25]. To our knowledge, however, this is the first attempt to use multiple interactive layers of DNA to obtain fine grained control of particle assembly.

To study the DNA functionalized nanoparticles, we use the model of Knorowski et al. [26], later extended by Li et al. [27] for double stranded DNA chains. This model, shown in Figure 1-b and which we will henceforth refer to as 'fine-grained' to distinguish it from a coarser model described later, lumps the interactions of multiple atoms at a level that retains chemical specificity, i.e., it can be connected directly to experimental parameters, while allowing relatively high computational performance in simulations.

The DNA functionalized nanoparticles are modeled as having a rigid-body repulsive core, connected to strands of double stranded DNA spacers, which are also repulsive. At the end of each chain, there is a single stranded DNA segment, where interactive beads are surrounded by repulsive flanking beads that connect to the interactive bead and the corresponding spacer bead. These flanking beads guarantee that chain-chain interactions only occur in a parallel manner, precluding multiple nucleotide associations and mimicking DNA hybridization. Each interactive bead is attractive to only one other type. Here we use two pairs of such selective beads, shown in Figure 1-b as a light blue and dark blue pair and a light orange and dark orange pair.

The fine-grained DNA model was studied using molecular dynamics (MD) with the Large-scale Atomic/Molecular Massively Parallel Simulator package (LAMMPS) [28]. All simulations were specified using Lennard Jones units. We follow the level of coarse graining in Knorowski. et al. [26] and Li et al [27] where $\sigma = 1$ corresponds to approximately $2nm$ in a real system. The core diameter of our particles is $10\sigma$, corresponding to $\sim 20nm$, composed of a rigid body of atoms as imposed by the *fix rigid* command. All repulsive beads interact through the Weeks-Chandler-Andersen (WCA) potential [29], with double stranded spacers having $\sigma = 1$ and a cutoff of $2^{1/6}\sigma$. Flanking beads have $\sigma = 0.6$, with a correspondingly scaled cutoff.

For repulsive interactions, the energy level is set to $\epsilon_{rb} = 1$. Attractive beads interact through a shifted Lennard Jones potential with $\sigma = 0.6$ and $\epsilon_{ab} = 10$. As discussed in [26], the system is not sensitive to the specific choice of $\epsilon_{ab}$ as long as $\epsilon_{ab} \gg \epsilon_{rb}$. The cutoff for the attractive Lennard Jones interactions was set to 3.0.

All bonded interactions are applied through a harmonic potential:

$$V_{bond} = \frac{1}{2}k(r - r_0)^2$$

Where $k = 330$ for all bonds and $r_0 = 0.84$ for bonds except for flanker-spacer bonds, which have a $r_0 = 1.19$.

We also apply a harmonic angle potential that aligns neighboring interactive beads in a chain. Similarly, we apply this potential to align each interactive bead to its two flanking beads.

$$V_{angle} = \frac{1}{2}k(\theta - \pi)^2$$

For these two potentials $k = 120$. To model the stiffness of double stranded DNA, we follow Li et al [27] by applying a similar harmonic angle potential to the double stranded spacer beads, with $k = 10$.

**Selection of Design Space**

Using this model, we can select a specific design space for our exploration as depicted in Figure 1-b. We fix the side chain motifs and particle size ($20nm$), selecting to modify only the following parameters: long DNA length, short DNA length, long DNA chain grafting density, and short DNA chain grafting density. In all designs, the parameters for particles A and B are the same, differing only by the (fixed) side chain motifs.

For the long DNA, we vary the length between 2 and 15 spacer beads. For the short DNA we vary the length between 1 spacer bead and the length of the long DNA in the specific design. For both long and short DNA, we vary the grafting density of chains between 5 and 70 chains, with the condition of a maximum of 100 chains on the particle. These choices provide a suitably large space (~300,000 total combinations) that allow us to fine-tune the interactions between the particles until we achieve designs that capture key features of the KM potential in such a way that they favor the formation of the double gyroid phase.

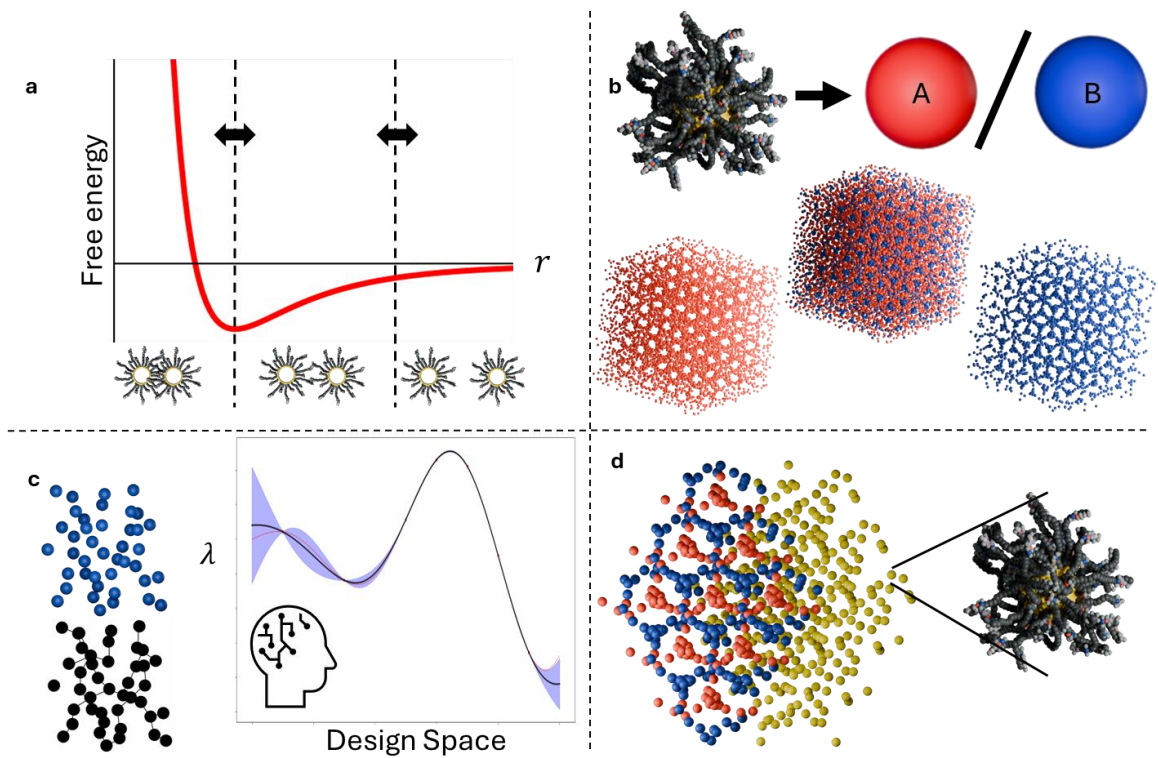

Figure 2: Key stages of Methodological framework. (a). Hamiltonian Replica Exchange Umbrella Sampling free energy calculations are performed of our fine-grained model to compute the average interactions of specific proposed designs. (b). The resulting Potentials of Mean Force coarse grain the DNA functionalized nanoparticles into single beads which are used to effectively model the self-assembly. (c). The self-assembled structures are scored by a graph-based order parameter, which measures how close the minority-component domain structure is to a perfect double gyroid. This order parameter is then fed into a Gaussian Process Regression machine learning model that provides predictions for the whole design space. (d). Designs predicted to form target phase are validated via interfacial pinning calculations of the fine-grained model. Only the cores of the minority component are shown, colored to make the gyroid structure apparent and distinguish it from the isotropic phase.

**Potential of Mean Force Calculations**

While we could evaluate the self-assembly of each possible design directly using the fine-grained model of DNA functionalized nanoparticles described above, this is not computationally practical with commonly available resources. The system sizes necessary to study self-assembly would require the use of millions of atoms, with the large number of degrees of freedom in the system making assembly process sluggish using direct molecular dynamics simulations. However, to resolve the relevant sections of the design space, many such simulations must be performed; the key is to try to make such a sampling selective rather than exhaustive.

One approach to implement the sought-after sampling selectivity is to use coarser-grained models that can be more extensively sampled as 'filters' to identify promising candidate designs. Here we implement a multiscale approach with two levels of coarse graining: One by mapping the fine-grained model into a coarse-grained particle model having single-pair interaction potentials, and the other by mapping the latter model's data into a machine learning correlator.

We hence first use free energy calculations to further coarse grain the interactions between fine-grained particles, reducing the number of degrees of freedom dramatically, as depicted in Figure 2-a. For each design under study we run three simulations, one for each interactive pair of nanoparticle types, where we compute the potential of mean force (PMF) between the particles. This calculation is performed using umbrella sampling [30]. We set up 39 evenly spaced windows between a distance of $12\sigma$ (near contact) and $50\sigma$ (where particles do not interact) and use a biasing harmonic spring to control the distance between the center of mass of each particle in the different windows. We then collect histograms of separation distances by sampling each window. These statistics can then be unbiased and patched together using the Weighted Histogram Analysis Method (WHAM) [31], to obtain the free energy of the system, i.e., the PMF, as a function of distance.

In our fine-grained system, at distances where particles are near the well of the PMF, it is likely that their binding will be strong enough to slow conformational changes in the DNA functionalized particles, negatively affecting ergodic sampling. To ameliorate this problem, we implemented a Hamiltonian Replica Exchange (HREX) [32] algorithm in LAMMPS by modifying the capabilities of the *fix temper* command, changing the order parameter for the exchanges to distance instead of temperature following Prajwal et. al. [33]. Code for this modification is available at [34]. In the HREX approach, simulations can switch between adjacent windows according to a Metropolis acceptance criterion. This allows structures which are "stuck" in conformational space to move to other windows with varying degrees of inter-particle coupling (being weaker at longer distances), thus improving ergodic sampling.

In the HREX calculations, we use a uniform spring constant of 6 in all windows. Simulations are carried out using the NVT ensemble at a temperature of 1.0 in LAMMPS, using a timestep of 0.005, a damping constant of 0.5 and running each simulation for 10,000,000 timesteps. Attempts to switch between umbrellas occur every 2,000 steps.

**Coarse grained self-assembly simulations**

Once the two-particle PMFs have been computed for a given design (for AA/BB and AB pairs), we use them to evaluate the many-particle self-assembled structure obtained through a Brownian Dynamics simulation (BD), where each "pseudo" atom represents a full nanoparticle which now interacts through the corresponding PMF as shown in Figure 2-b. This reduces the number of degrees of freedom by three orders of magnitude and allows reasonable system sizes to be explored. For each design, we run a simulation with 10,000 coarse grained atoms, at a ratio of 67% type A and 33% type B particles, which is the composition expected to favor the double gyroid. These simulations are performed in LAMMPS using a Langevin thermostat at a temperature of 1.0 with a damping constant of 1.0 and a Berendsen barostat at a pressure of 1.0 with a damping constant of 10.0. The timestep for these simulations was 0.01 and the system was run for 20,000,000 steps. The configuration at the end of the simulation is saved and used to evaluate the resulting self-assembled structure of the minority component.

**Double Gyroid Order Parameter**

To evaluate whether the system converged to a gyroid morphology, we need to use a proper order parameter that can distinguish the gyroid from other phases encountered when sampling the design space. Ideally, this order parameter would give a quantitative metric of how close the system was to a reference gyroid structure, which is essential to inform and guide any training algorithm. For this purpose, we chose an order parameter based on graph distance metrics as depicted in Figure 2-c. To compute it, we first sample configurations after the structure has converged and compute their pair distribution function for the minority component. The first minima sets a cutoff distance for nearest neighbors. We then cluster the minority component based on this cutoff and turn the largest cluster into graphs, where each atom corresponds to a node and the connectivity between nodes is set by their neighbors. We can then use a graph distance algorithm to compare them to the graph of a baseline double gyroid, computed by simulating the KM potential directly.

Many distance metrics for graphs exist with a particularly appealing choice being Portrait Network Divergence (PND) [35]. This metric compares the portrait network of each graph, which measures how a graph grows as it moves away from a given node. Since in our graph each node corresponds to a particle, the portrait network would correspond to a measure of how the number of neighbors of neighbors grow at various layers away from each particle. This gives it an intuitive connection to the physical connectivity of the structure. The PND of the two graphs is then a comparison of how differently they grow. Conveniently, it is a normalized metric, can be used for graphs of different sizes, and PND values near 0.0 correspond to very close graphs and values near 1.0 corresponds to very distant graphs. We use the python *netrd* implementation of PND [36]. For this work, we use $\lambda = 1 - PND$ as our order parameter, so that more structured phases lead to larger values of the $\lambda$.

In addition to the use of $\lambda$ as a continuous measure of double gyroid structural proximity, we identify fully assembled double gyroids both visually and by computing their structure factor $S(k)$, which can be computed through the equation:

$$S(\boldsymbol{k}) = \frac{1}{N}\left\langle \left|\sum_{i=1}^{N} \cos(\boldsymbol{k} \cdot \boldsymbol{r}_i)\right|^2 + \left|\sum_{i=1}^{N} \sin(\boldsymbol{k} \cdot \boldsymbol{r}_i)\right|^2 \right\rangle$$

where $\boldsymbol{r}_i$ are the positions of the particles and $\boldsymbol{k}$ are the wavevectors, which are chosen from the set of $\frac{2\pi}{L}(n_x, n_y, n_z)$ where $L$ is the side length of the simulation box and $n_{x/y/z}$ are natural numbers. The structure factor for a double gyroid has peaks at relative positions: $|k|/|k^*| = \sqrt{6}, \sqrt{8}, \sqrt{16}, \sqrt{20}, \sqrt{22}, \sqrt{24}, \sqrt{26}$ [37].

**Gaussian Process Regression and Exploitation Active Learning**

As the last stage of our multiscale framework, the order parameter data from above coarse-grained model simulations are fed into a machine learning algorithm, which can predict $\lambda$ for all other designs; i.e., enact an exhaustive sampling. We chose a Gaussian Process Regression (GPR) [38] algorithm for this task, as it is well suited to handling relatively small datasets. We use a standard scaled RBF Kernel with constant mean via the GPyTorch implementation [39] of the algorithm.

After using the GPR model with an initial data set, we have a set of predictions for each design, along with the uncertainty of each prediction. This data can allow us to follow an active learning optimization technique to choose the next set of designs to evaluate. Usually, one can use either an exploration or an exploitation strategy (or a balanced mix of both) [40]. An exploration strategy seeks to fill out sampling gaps over the whole parameter space, at each step choosing the set of predictors with the highest uncertainty. In contrast, an exploitation strategy attempts to focus the sampling on the area of the parameter space most likely contain the sought-after output, by choosing the next few designs with the best (smallest-value) predictors for the order parameters.

Since our goal is only to find designs (or specific regions in parameter space) that lead to the double gyroid phase and not a thorough exploration of all possible structures, we follow an exploitation strategy. The next set of suggested designs are then evaluated to compute their

$\lambda$, and the GPR algorithm is run again with an updated dataset. This loop is repeated until a reasonable number of double gyroid designs have been found, effectively mapping out at least one "double gyroid peak" in parameter space.

**Interfacial Pinning Simulations**

Once designs that form the double gyroid have been identified, we conduct a consistency check to verify whether the PMF-based coarse grained model correctly predicted self-assembly of the fine-grained model. For this, we choose one characteristic design from the gyroid basin and evaluate it at the fine-grained scale. Ideally, one would study the spontaneous self-assembly of this system from a disordered state but, as discussed earlier, this process is extremely slow and computationally expensive to be practical given our resources.

Instead, we can probe the stability of the phase with respect to that of the disordered state using the interfacial pinning method [41-42], as depicted in Figure 2-d where we show only the cores of the minority component for clarity. In this method, we set-up a two-phase configuration of the particles, one phase being the double gyroid (with particles colored to distinguish the networks) and the other the isotropic phase (with an even color). Then we apply a harmonic field of strength $\kappa$ over fluctuations on a suitable system parameter $Q$ that, having distinct values for each phase, is given an intermediate "anchoring" value $a$, to bias the system to maintain the two-phase configuration. The Gibbs free energy difference between the phases can then be estimated by the average biasing force exerted on the system, following the equation:

$$\Delta G \approx -\frac{\kappa \Delta Q}{N}[\langle Q \rangle - a]$$

where $\Delta Q$ refers to the difference between a chosen order parameter of the phases under consideration (here isotropic and double gyroid), $\langle Q \rangle$ is the average value computed during the pinning simulation and $N$ is the number of molecules under study. In this work, we chose the total potential energy of the system as our pinning order parameter $Q$. It is a convenient choice as it is easily available from independent simulations of pure double gyroid and pure isotropic configurations and is an inexpensive order parameter to compute, which is convenient given the large system size of the pinning simulations.

The pinning was implemented using the PLUMED [43-45] library combined with LAMMPS. We use *MOVING RESTRAINT* feature to set-up the harmonic biasing force on the system, slowly increasing the spring constant from a very small value until the desired strength for the bias is achieved. After equilibration, we can sample the potential energy of the system to compute the Gibbs Free Energy difference between the two phases, whose sign indicates which of them is more stable.

## Results and Discussion

**Potential of Mean Force Calculations**

As explained in the Methods Section, we began the exploration of the design space by randomly selecting 10 designs and calculating their PMF. This process was repeated until a total of 100 designs were explicitly evaluated and the design space predictions had reasonably converged. Three characteristic example designs chosen from the final dataset are shown in Figure 3, which shows the resulting PMF along with the self-assembled structures, showing only the minority component for clarity.

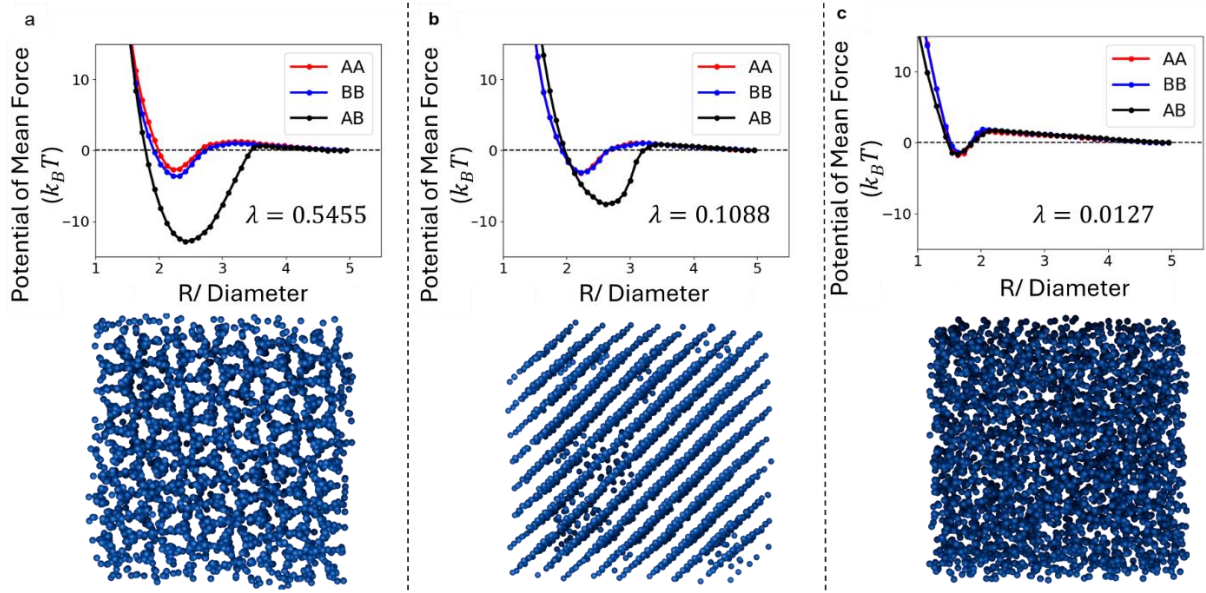

Figure 3: Results of PMF calculations and coarse-grained configurations for three example designs. For clarity, only the minority component is shown. (a): Results for a design which assembles into a double gyroid. Note the positive non-additivity in the potential. (b). Results for a design assembling into a perforated lamellar phase. (c). Results for a design which does not undergo self-assembly. Note decreasing value of $\lambda$ from I to II to III.

The first design embodies some of the key features of the KM model, with AA and BB interactions being weaker, but closer, than AB interactions. It results in a double gyroid phase, with a correspondingly large order parameter value of $\lambda = 0.5455$. The second design also resulted in a non-additive PMF, but self-assembles into a perforated lamellar phase instead. This results in a low value of $\lambda = 0.1088$. The third design has relatively small energy wells, leading to weak attraction. Additionally, the interactions between the AB pair are very similar to the AA/BB pair. This lack of preference for any specific pair leads to a random structure, which has a correspondingly vanishingly small value of $\lambda = 0.0127$.

From these results we can see that our order parameter follows the expected behavior, with smaller values of $\lambda$ corresponding to structures farther away from the double gyroid. Note that $\lambda \sim 0.5 - 0.6$ is characteristic of most successful designs and seems to be an upper limit to the values computed by our order parameter. Reaching a value lower than $1.0$ for $\lambda$ is to be expected, given small fluctuations and imperfections in the double gyroid structure as a function of time. It is also noteworthy that there is some variation in the values for the order parameter with both false negatives (double gyroids with $\lambda \sim 0.3$) and false positives (non-double gyroid with $\lambda \sim 0.6$). This, however, does not seem to impede the convergence of the machine learning algorithm significantly, as shown below.

Interestingly, it is not enough to satisfy the conditions of the KM model to form a double gyroid, since only a subset of designs that showed these characteristics were successful. During their exploration of the potential, Kumar and Molinero found the gyroid phase was formed by only a certain range of ratios of energy well location and depth. It is likely that the same principle applies here, and only a subset of ratios will be successful. Additionally, the specific shape of the PMF also matters for assembly, as for example, having too strongly repulsive cores can preclude assembly into mesophases, as explored in [22].

One common feature in many of our calculated PMF's, that is absent in the simplistic Kumar Molinero potential that exhibits an attractive long-distance tail, is the presence of a small repulsive bump at intermediate interparticle distances which dies off at longer distances (than any attraction. This is likely due to steric repulsion from the long chains overlapping before they get to the distances where they can successfully hybridize. Another difference from the classical KM model is the well depths, which here can be up to $10 k_B T$ deep, giving far stronger binding than those studied in KM, which were around $4k_BT$. This difference does not seem to preclude assembly, albeit may hamper reversible binding and thus slow down equilibration .

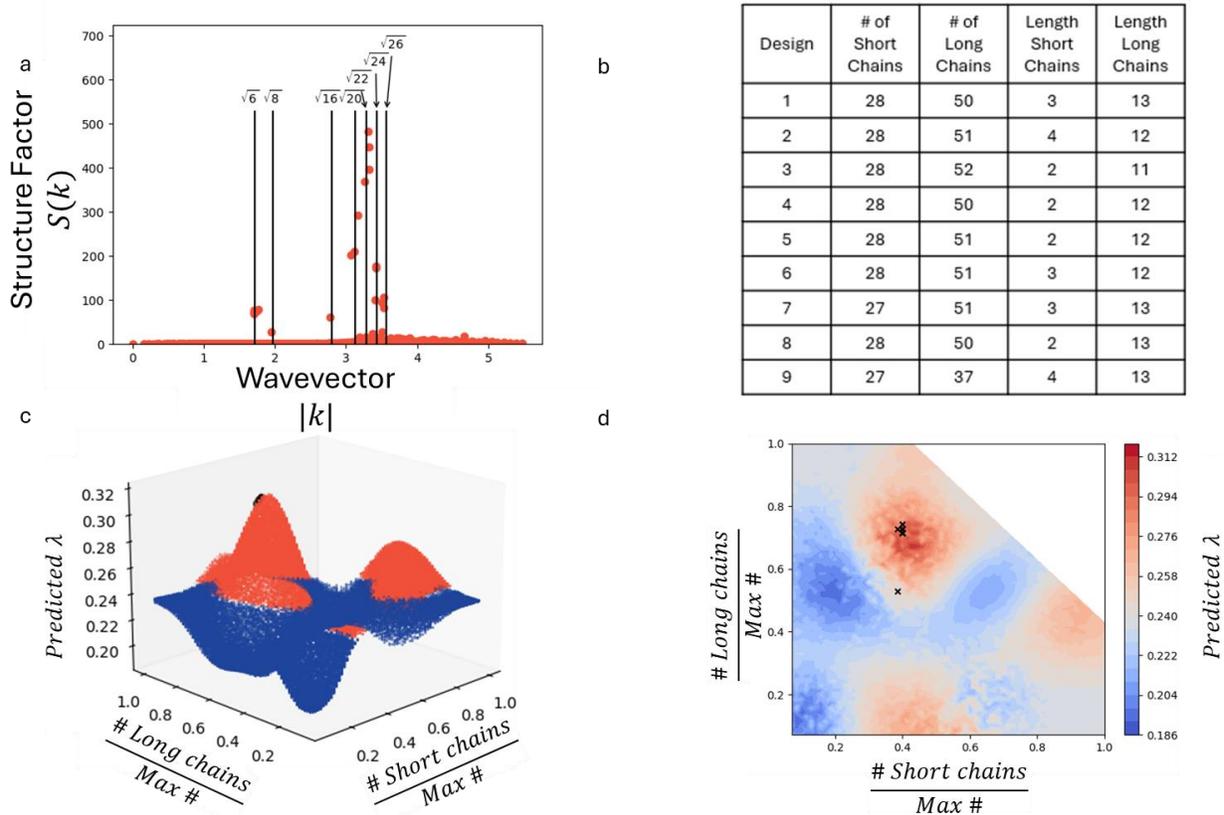

Figure 4: Results of the converged GPR machine learning model of the system, along with a table of successful double gyroid designs. Areas of high $\lambda$ are shown in red and low $\lambda$ are shown in blue. Successful designs are marked in black. (a) Structure factor of example successful double gyroid. (b) Table of specific successful designs for the double gyroid. (c) Predictions as a function of the grafting densities. (d) Top down view of design space as a function of grafting densities.

**Exploration of Parameter Space**

Figure 4 shows the results of the complete exploration of the parameter space after using our exploitation strategy. It represents data from 100 simulations, which is 0.03% of the total design space. As the design space is four-dimensional, we focus on two variables for ease of visualization. Variables are normalized by their maximum allowable value. In these plots, the regions of high $\lambda$ (relative to the average value of the predictions) are shown in red and regions of small $\lambda$ in blue. Successful double gyroid designs (as per the CG model) are marked in black.

In Figure 4-a, we show a characteristic structure factor for a successful double gyroid where we can see the presence of the expected pattern of diffraction peaks. In Figure 4-b we show the specific designs that led to successful double gyroids. It can be clearly seen that all these designs are near each other in terms of the design parameter values. In Figure 4-c the predicted values of $\lambda$ for the whole design space are plotted as a function of the grafting densities. We can observe the presence of several well-defined peaks and basins. All the successful double gyroid designs are clustered on the main tallest peak, showing that the machine learning algorithm successfully identified a region of the design space conducive to the gyroid phase. For clarity, in Figure4-d this design space is plotted as a contour map. We can see that almost all of the double gyroids fall on the top of the main peak, with only one of the designs falling on the boundary of this peak. No double gyroids were found on the other shorter peaks. These peaks may be associated with regions where the double gyroid may exhibit some weak level of metastability.

It is important to note that the GPR model condenses the range of predicted $\lambda$ to $\sim[0.15-0.35]$, when the actual data varies from $\sim[0.01-0.6]$. This is likely a result of the GPR pushing the results towards the mean of the data. This is not a limitation, however, as the goal is to find the location of the peaks, not to have accurate predictions for $\lambda$.

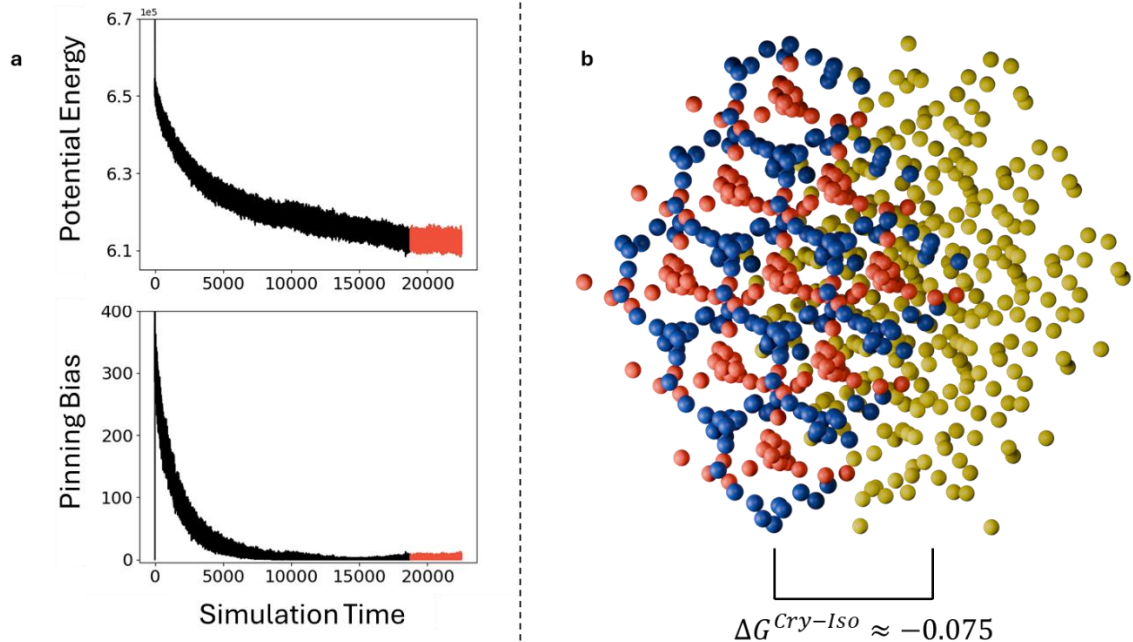

Figure 5: Results of interfacial pinning simulations for design #1, chosen as a representative sample of the double gyroid basin. (a). Results of the potential energy and pinning bias as a function of time. Highlighted in red is the period when sampling has reached equilibration. (b) Example of converged configuration of the interfacial pinning simulation showing only particle cores. Cores colored yellow for the isotropic region and blue and red for the double gyroid region, where the two colors distinguish the two networks.

Since we know the experimental relationship between the number of spacer beads and the length of the double stranded DNA (each bead corresponds to approximately $2nm$ of dsDNA) this can serve as a direct guide to the experimental synthesis of these particles.

**Interfacial Pinning Calculations**

We chose design #1 as a random sample from the main gyroid basin to analyze its stability in detail. For this system, the anchor energy for the harmonic bias was $a = 615,536$, (in LJ units) computed from the average potential energies of the gyroid and isotropic phases. The bias was initially set at $a = 735,420$, which was the initial potential energy of the configuration, and a weak force constant of $\kappa = 1 \times 10^{-7}$. These choices were then gradually changed over the first few steps of the simulation to the target anchor value and

$\kappa = 5 \times 10^{-7}$. This procedure helped prevent numerical instability when applying the bias. The simulations were conducted in the NPT ensemble at a temperature of 1.5 (in LJ units) and a pressure of 0.0. This temperature was chosen to guarantee mobility in the system [26] and, given that the dominant interactions in the system are attractive, the system should have a weak dependence on pressure.

The results from this simulation are shown in Figure 5. In Figure 5-a we show the convergence of the potential energy and pinning bias of the pinned system as a function of simulation time. The final average potential energy of the pinned system was $\langle Q \rangle = 612{,}572.5$. This resulted in a value for the Gibbs Free Energy of $\Delta G^{Cry-Iso} = -0.075$, showing that the gyroid phase is more stable than the isotropic at the simulation conditions. This, combined with the fact that the phase spontaneously self-assembles at the coarse-grained level, is strong evidence that the phase should be experimentally attainable with DNA-grafted particles consistent with design #1.

## Conclusion and Outlook

The problem of connecting a desired self-assembled structure with appropriate building blocks is an important open problem in soft matter physics. We approach this problem using DNA functionalized nanoparticles as a platform to enact interaction control. The design of these nanoparticles is informed by existing knowledge of effective pair potentials that lead to a desired phase, here the double gyroid. However, even using this potential as a guide, it is a non-trivial task to identify realistic building blocks that could give rise to such effective pair potentials, and any proper description of such building blocks will likely necessitate computationally demanding simulations and the search of a large design parameter space.

We explore the parameter space of the particles by first using free energy calculations to coarsen a fine-grained model of the DNA nanoparticles by finding their effective pair potential. This potential is then used to explore the self-assembly of the designs of interest in a computationally efficient manner. Combining this with a Gaussian Process Regression machine learning model allowed us to map out the relevant regions of the design space that lead to a double gyroid. We additionally found specific designs that converged to the desired structure and verified their stability through further free energy calculations of the fine-grained model. These designs of interest can be directly connected to experimental realizable nanoparticle constructs, thus giving a direct recipe for their synthesis.

While the framework put forward in this work proved effective in the systems studied, it does have several limitations. For instance, the model scope is limited by the validity and accuracy of the fine-grained description adopted for the particles under study. For more complex particle constructs, this might not be readily available. It is also possible for available models to be too fine grained (such at the all atom level) precluding effective evaluation of the potential of mean force. Another important source of error can come from the choice of order parameter used to describe the proximity to the desired phase. This can increase the number of false positives and false negatives in the structural order parameter predicted by the machine learning model. In our case, the machine learning model seemed to be capable to adjust for these errors, but it is important to consider them when evaluating the quality of the data and the sample sizes to explore. Finally, even though the coarse-grained approach is fast, its timely convergence is not guaranteed with the typical durations of MD runs, and its results must always be validated in some way by the finer description, to make sure the extra degrees of freedom do not affect the assembly.

While DNA based technology for nano-scale assembly is a well explored field (see [15-18,46]), our multiscale approach to parameter space exploration, that combines coarse-graining free energy calculations and machine learning models, is a strategy with a potentially broad range of applicability. The PMF calculation can be used to coarse grain many building blocks and can be extended to include multi-body interactions if needed. Additionally, our graph distance-based order parameter could prove effective in quantifying the quality of other types of network structures. This underscores the potential of our coarse-graining approach to quickly and efficiently explore complex parameter spaces and narrow down conditions that can produce materials with targeted properties for applications in photonics, catalysis, biomedicine and other fields.

## Acknowledgements

The authors would like to thank the NSF for support through NSF Award No. CHE-2101829.